# Control of hot-carrier relaxation time in Au-Ag thin films through alloying

Sarvenaz Memarzadeh, Kevin J. Palm, Thomas E. Murphy, Marina S. Leite, and Jeremy N. Munday*

S. Memarzadeh
Department of Electrical and Computer Engineering, University of Maryland, College Park, Maryland 20742, USA
Institute for Research in Electronics and Applied Physics, University of Maryland, College Park, Maryland 20742, USA
K. J. Palm
Department of Physics, University of Maryland, College Park, MD 20742, USA Institute for Research in Electronics and Applied Physics, University of Maryland, College Park, MD 20742, USA
Prof. T. E. Murphy
Department of Electrical and Computer Engineering, University of Maryland, College Park, Maryland 20742, USA
Institute for Research in Electronics and Applied Physics, University of Maryland, College Park, Maryland 20742, USA
Prof. M. S. Leite
Department of Materials Science and Engineering, University of California, Davis, CA 95616, USA
Prof. J. N. Munday
Department of Electrical and Computer Engineering, University of California, Davis, CA 95616, USA
Department of Electrical and Computer Engineering, University of Maryland, College Park, Maryland 20742, USA
Email Address: jnmunday@ucdavis.edu

## Abstract:

The plasmon resonance of a structure is primarily dictated by its optical properties and geometry, which can be modified to enable hot-carrier photodetectors with superior performance. Recently, metal-alloys have played a prominent role in tuning the resonance of plasmonic structures through chemical composition engineering. However, it has been unclear how alloying modifies the time dynamics of generated hot-carriers. In this work, we elucidate the role of chemical composition on the relaxation time of hot-carriers for the archetypal $Au_x Ag_{1-x}$ thin film system. Through time-resolved optical spectroscopy measurements in the visible wavelength range, we measure composition-dependent relaxation times that vary up to 8x for constant pump fluency. Surprisingly, we find that the addition of 2% of Ag into Au films can increase the hot carrier lifetime by approximately 35% under fixed fluence, as a result of a decrease in optical loss. Further, the relaxation time is found to be inversely proportional to the imaginary part of the permittivity. Our results indicate that alloying is a promising approach to effectively control hot-carrier relaxation time in metals.

## Introduction

Pure metals, such as gold (Au) and silver (Ag), have long been the most commonly used plasmonic materials due to their high electron densities and desirable optical and chemical properties. However, when using pure metals, applications are limited to a narrow range of optical frequencies stemming from the inherent resonances of the metals. Alloying these metals together presents a promising alternative by allowing the opportunity to tune the plasmonic resonances without altering the geometry of the system. The optical properties of the Au-Ag alloys can be tailored throughout the visible spectrum by modifying the atomic ratio of the two metals [1, 2, 3, 4].

Additionally, by varying the alloys' chemical composition, one can modify the electronic band structure, which results in inter-band transitions over different incident photon energies. It was recently reported that as the concentration of Au increases in Au-Ag alloyed films, the position of the d-band shifts closer to the Fermi level[5]. This reduces the energy gap for inter-band transitions, leading to transitions occurring with lower incident photon energies. Similar modification of the threshold of the inter-band transitions has also been studied in other types of materials such as metal nitrides [6], semiconductors [7], and transition metal dichalcogenides [8]. The resonance tunability and band structure engineering of alloys proves useful in a variety of applications including superabsorbers [9], imaging probes in biomolecular studies [10, 11], implant devices [12], catalysis [13, 14, 15], photovoltaics [16, 17], and hydrogen sensing [18, 19, 20, 21].

Many of the aforementioned applications rely on significant light absorption within the films or nanostructures. One common approach for absorption enhancement is through coupling the incident photons into surface plasmons, i.e. coherent oscillations of free electrons at the metal-dielectric interface. This process results in the generation of highly energetic non-thermal carriers, also known as hot-carriers. Particularly, hot-carriers are generated after nonradiative decay of the localized or propagating surface plasmons through either direct or phonon-assisted intra-band transitions [22, 23]. Once these carriers are excited, they thermalize to create a population of electrons that can be described as a Fermi-Dirac distribution at an elevated temperature. They start to equilibrate with the lattice temperature via a series of scattering processes including the electron-phonon and phonon-phonon scatterings [24, 25]. These highly energetic carriers have been utilized in applications such as water splitting [26], artificial photosynthesis [27], medical therapy [28], and drug delivery [29]. However, efficient generation and extraction of these carriers depends on the choice of material, and their corresponding hot-carrier relaxation time. In particular, understanding of the hot-carrier relaxation time plays a significant role in modulation speed [30], power conversion efficiency enhancement [31, 32], determining the hot-electron flux [33, 34], and nanoscale photothermal heat control [35]. Thus, due to the broad spectral tunability associated with devices exploiting hot-carrier physics, their temporal study in planar Au-Ag structures would benefit a variety of applications.

Here, we focus on Au-based hot-carrier devices due to their chemical stability and incorporate different ratios of Ag to create Au-Ag alloys. We use ultrafast pump-probe optical spectroscopy to measure the hot-carrier relaxation time. The pump wavelength is nominally set to 700 nm wavelength (1.77 eV) to ensure that the relaxation time is due to intra-band transitions rather than inter-band ones (2.4 eV in Au and 4.0 eV in Ag) [36, 5]. We further employ the Kretschmann geometry to couple into the propagating surface plasmon mode, which has the added benefit of increasing the measurement sensitivity as a result of increased photon absorption. To determine the hot-carrier lifetime, we use a free-electron model and convert the differential reflectivity measurements to the corresponding elevated electron temperature [37]. Our results show that the hot-carrier relaxation time depends upon the Ag mole fractions. We further find that the lifetime is inversely proportional to the imaginary part of the permittivity for different Au-Ag alloys. Finally, considering the pure Au film as the baseline of the lifetime measurements, we observe that the slight addition of Ag (2%) can increase the hot-carrier relaxation time, while higher fractions of Ag (i.e. 35% and 75%) yield smaller lifetimes.

## Results and discussion

To study the effect of alloying on the hot-carrier relaxation time, we fabricated four samples with different Au-Ag compositions using a co-sputtering deposition procedure. We change the alloys' composition by varying the applied voltage on the Au and Ag targets during each deposition. The chemical composition of the alloyed samples is determined with energy-dispersive X-ray spectroscopy (EDS) (see Fig. S1). Subsequently, we measure the optical properties of our samples with spectroscopic ellipsometry in the wavelength range from 200 nm to 1000 nm. We use a Drude-Lorentz model including two Drude and one Lorentz terms to _t the ellipsometry data. The modelled permittivity is shown in Figure 1(a) and 1(b) for our fabricated samples with the subsequent fitting parameters (see Table S2 and S5). We also compute the surface plasmon polariton (SPP) quality factor, $Q_{spp}(\omega) = \frac{\epsilon_r^2(\omega)}{\epsilon_i(\omega)}$ [38], for the different Au-Ag alloys (Fig. 1(c)). In general, many experimental factors such as the chamber pressure, substrate temperature, deposition rate, etc. can affect the films' quality factors due to the change in the dielectric functions [39]. Our experiments keep all of these other factors the same, thus isolating the effects of changing the alloy composition. At 700 nm, the wavelength used for our pump-probe measurements, the 100% Au and 98% Au samples show a higher $Q_{spp}$ when compared to the other alloys, predominantly due to the lower $\epsilon_i(\omega)$. The dielectric functions can also be affected by a disordered mixture of Au and Ag at a certain molar combination, leading to the reduction of electron scattering and plasma frequency [5, 40]. Additionally, it has been shown that the co-sputtering of a small amount of metal suppresses the island growth, leading to a film with low optical and electrical losses [16].

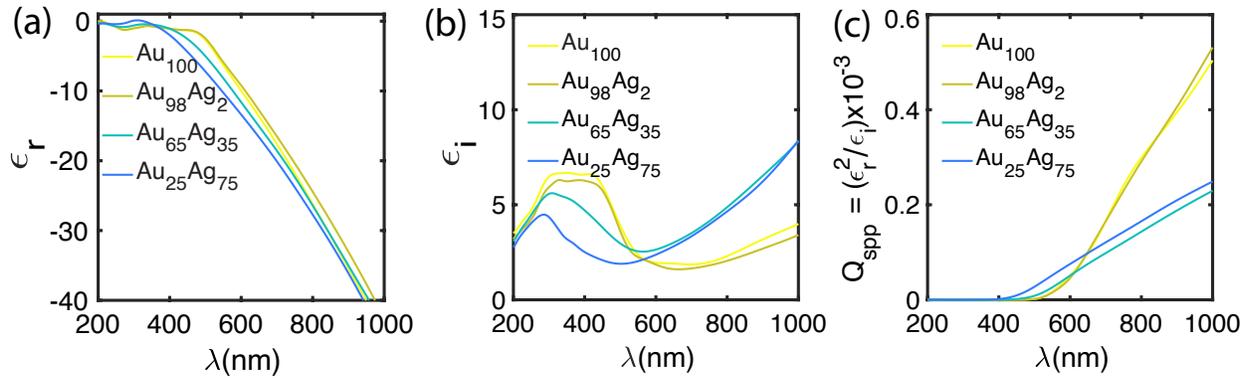

Figure 1: Optical response of Au$_x$ Ag$_{1-x}$ alloys. Measured (a) real and (b) imaginary parts of the permittivity, and (c) computed quality factor of the propagating surface plasmon.

Before measuring the relaxation dynamics of the excited hot-carriers, we measured the propagating surface plasmon mode using the Kretschmann configuration [41]. Figure 2a shows the experimental results of the reflection measurements for all four samples as a function of incident angle near the plasmon coupling angle for incident wavelength from 680 nm to 740 nm with 5 nm spectral bandwidth.

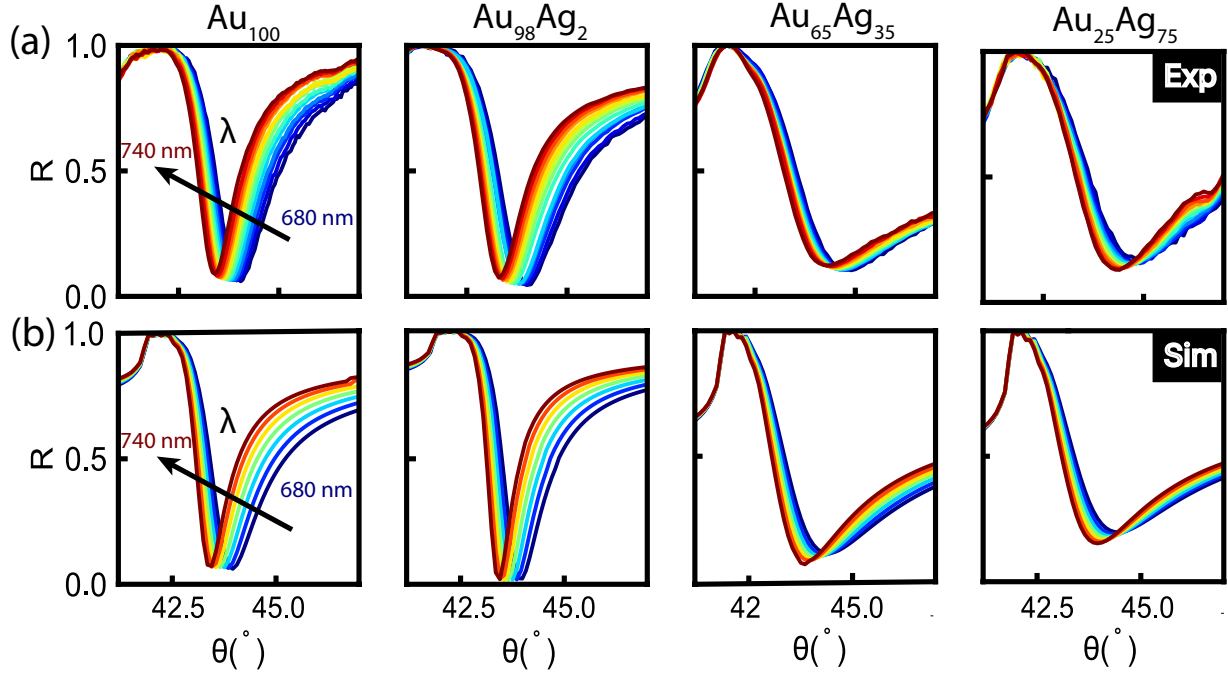

Figure 2: (a) Experimental and (b) simulated reflectivity for Au, $Au_{98}Ag_2$ , $Au_{65}Ag_{35}$ , and $Au_{25}Ag_{75}$ alloys under the p-polarized illumination with wavelengths ranging from 680 nm to 740 nm. For FDTD simulation, we used pulse illumination with 150 fs pulse width.

As expected, the surface plasmon resonance is sharper for samples with higher $Q_{spp}$ and broader for the samples with lower values. Our measured optical properties (Fig. 1) are used in the Finite Difference Time Domain (FDTD) numerical simulations and are in good agreement with our experimental results (Fig. 2b).

The non-equilibrium hot-carrier dynamics of the alloys are investigated using degenerate ($\lambda_{pump}$ = $\lambda_{probe}$) time-resolved differential reflectivity measurements at the surface plasmon resonance angle. We use a Ti-Sapphire laser system with 700 nm wavelength and 80 MHz repetition rate to generate both the pump and probe beams. A fraction of the laser beam is split off to serve as the probe beam and the other portion is passed through a mechanical delay stage to set the time delay between the two beams. We use nearly co-linear pump and probe beams, which are adjusted to couple into the propagating surface plasmon mode but can also be spatially separated in the reflected field. The overlap of the beams is achieved using an off-axis parabolic mirror with a measured spot size of approximately 90 $\mu$m. Pump-probe measurements are conducted at the surface plasmon resonance angle under five different incident pump powers (i.e. 120 mW, 150 mW, 180 mW, 210 mW, and 240 mW) with a fixed probe power of 19.8 mW, as the results are indicated in Figure 3.

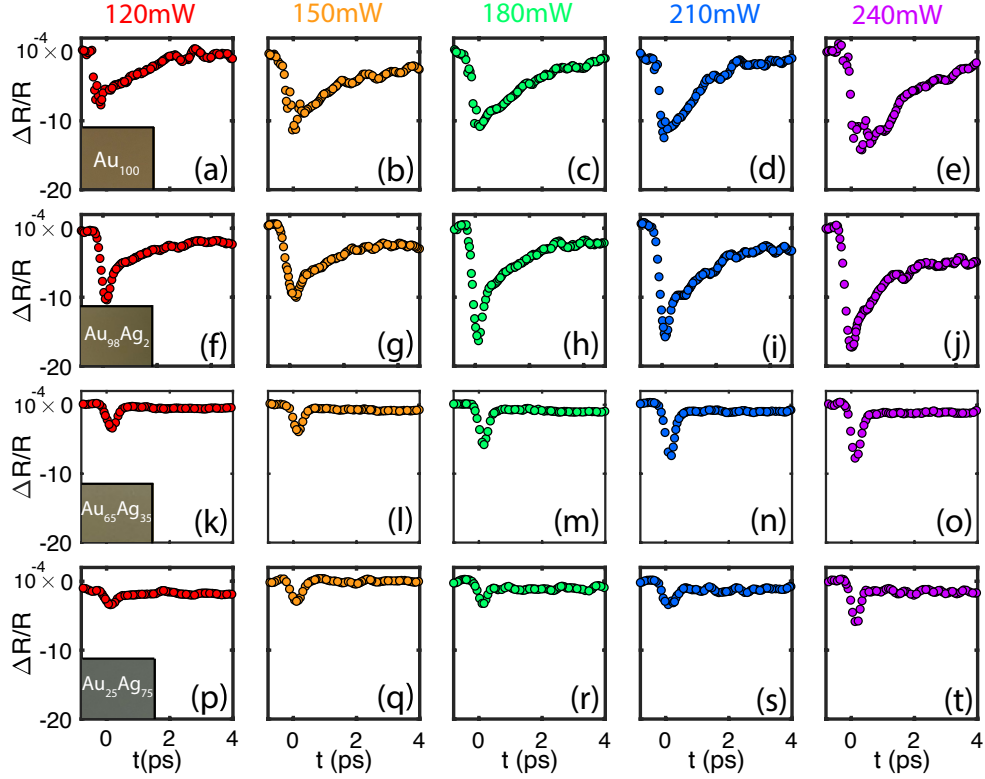

Figure 3: Differential reflectivity measurements for Au-Ag alloys with different chemical compositions. For each sample, the pump power is: 120 mW (a,f,k,p), 150 mW (b,g,l,q), 180 mW (c,h,m,r), 210 mW (d,i,n,s), and 240 mW (e,j,o,t). Insets are real-color photographs of the alloyed thin films.

As expected, in all cases, increasing the pump power produces a larger change in the transient reflectivity ($\Delta R/R$). Because the temporal pulse width employed here is longer than the electron-electron scattering time, on the order of 100 fs [42], the relaxation time for the optically excited hot-carriers is mostly governed by the electron-phonon relaxation time. Furthermore, at the resonance wavelength, the pump induced change in the reflectivity signal is negative, which is attributed to the intra-band transition and the excitation of hot-carriers for the fabricated alloys [42, 43].

## Theory and analysis

To find the excited hot-carriers relaxation time from the transient reflectivity measurements, we employ the combination of a free-electron model [44] and the modified two-temperature model [37]. In this model, the effect of the surface plasmons electric field profile is incorporated into the absorbed laser power density within the conventional two-temperature model which accounts for variation of the field in the vertical (surface normal) direction. This combination allows us to convert the pump-probe reflectivity signal to the relevant electron temperature (details in the supplementary information). This approach results in a more accurate theoretical modeling due to the nonlinear relationship between the reflectivity signal and the electron temperature. This theoretical model uses the optical parameters extracted from our ellipsometry measurements at room temperature for each alloy. Finally, best fits to the temperature converted reflectivity signals are computed by minimizing the Normalized Mean Squared Error (NMSE) of the hot-carrier relaxation time. Figure 4 shows the results of the temperature converted data (solid circles) and their corresponding best fits (solid lines) to the hot-carrier relaxation time of the alloyed Au-Ag films at pump powers of 120, 150, 180, 210, and 240 mW under the resonance condition, i.e. upon coupling to the surface plasmon mode.

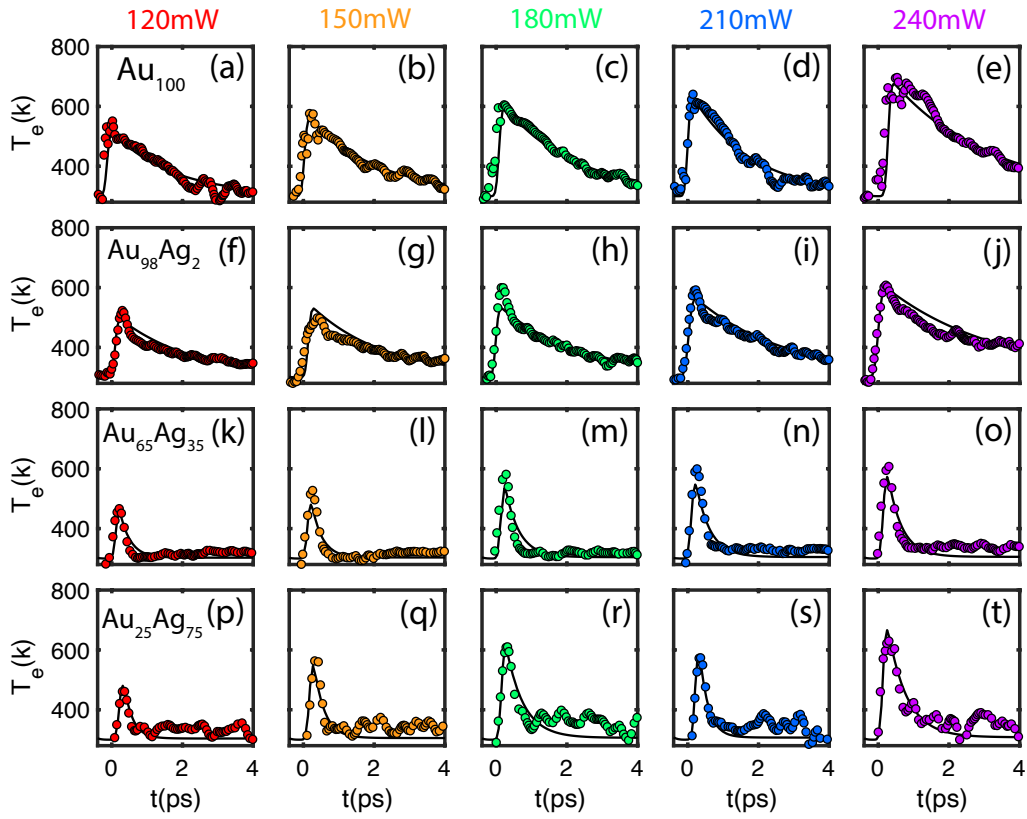

Figure 4: Temperature converted differential reflectivity measurements for Au100 , Au98 Ag2 , Au65 Ag35 , and Au25 Ag75 alloys under different incident pump powers of 120 mW (a,f,k,p), 150 mW (b,g,l,q), 180 mW (c,h,m,r), 210 mW (d,i,n,s), and 240 mW (e,j,o,t). The black solid lines in each plot show the best fits computed from a modified two-temperature model.

The pump-probe measurements also reveal an additional short decay component that only appears immediately after excitation for the 2% Ag composition. We attribute this decay component to electron-electron interactions, which are typically too fast to be detected in the pure Au. The plasmon dephasing time (i.e. the rate at which electrons collective oscillations cease) is longer in Ag as a result of different radiative or non-radiative plasmon damping mechanisms, and so the addition of Ag to the Au alloy may increase this decay component to a measurable amount in the 2% Ag alloy. For the higher Ag concentration alloys, this decay mechanism is not distinguishable from electron-phonon interactions based on our measurement sensitivity.

Analysis of the temperature converted differential reflectivity shows that the hot-carrier relaxation time($\tau$) of the $Au_{98}Ag_2$ sample is ~8 times larger than for $Au_{65}Ag_{35}$ and $Au_{25}Ag_{75}$ for a fixed laser fluence. Additionally, we find that the film with $Au_{98}Ag_2$ has the longest lifetime of any of the samples measured (3.20±0.15 ps with 240 mW pump power), even including pure Au. To further investigate this phenomenon, we consider the optical properties of each Au-Ag alloy at 700 nm pump wavelength and compare the result with our measured hot-carrier relaxation time and found $\tau$ to be inversely proportional to the imaginary part of the permittivity (Fig. 5).

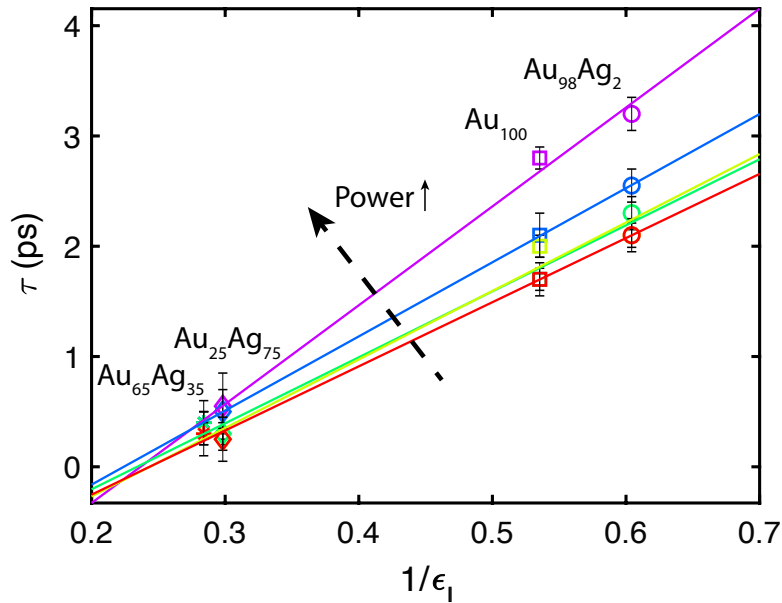

Figure 5: Hot-carrier relaxation time as a function of $\frac{1}{\epsilon_I}$, the inverse of the imaginary part of the permittivity, for $Au_x Ag_{1-x}$. The solid lines are the linear fit between the hot-carrier lifetime and $\frac{1}{\epsilon_I}$. The colors represent the range of the pump power between 120 mW (red) to 240 mW (purple).

Our results suggest that the addition of a small fraction (2%) of Ag to a Au film increases the hot-carrier lifetime. This is consistent with previous findings that showed particular Ag-Au alloys having higher $Q_{SPP}$ than pure metals [4] and that doping one metal with another can improve film quality and decrease optical loss [45, 16]. However, all alloyed films that we measured have similar surface roughnesses (see Fig. S3), suggesting that the decreased loss may come from changes in the band structure or other changes to the material rather than simply smoothing of the films. Because we are probing relaxation times >10s of fs, the main mechanism leading to the increase in the hot-carrier lifetime is likely a suppression of the electron-phonon scatterings, which could result for decreased lattice defects, grain boundaries, etc., but further work will be necessary to isolate the individual contributions. In addition, the lifetime is inversely proportional to $\epsilon_1$ and increases with pump power (see Fig. 5 for a comparison with all pump powers), which is in agreement with previously reported studies [46, 47]. These results show that the optical loss is an important and potentially controllable internal parameter compared to the other external factors, such as pump power. Our observation further opens a new route to alter the hot-carrier relaxation time for plasmonic applications through alloying.

## Conclusion

In summary, we measured the hot-carrier relaxation time of Au-Ag thin film alloys under visible excitation and found that adding a small fraction of Ag to Au increases the hot-carrier relaxation time. Our experimental results suggested that the relaxation time depends on the alloy's composition and is inversely proportional to $\epsilon_1$ . Surprisingly, some alloys can have loss factors that are less than their pure counterparts, which leads to improved hot-carrier performance. By comparing the relaxation time of the fabricated alloys with the pure Au sample, we determined that the measured relaxation time increases with slight addition of Ag and then drops significantly for alloys with higher Ag content. Overall, this work demonstrated that the relaxation time of hot-carriers can be engineered through alloying.

## Supporting Information

Supporting information is available upon request.

## Acknowledgements

This material is based on work supported by National Science Foundation CAREER Grant No. ECCS- 155450, Grant No. MMN-2016617, Grant No. MMN-1609414, and the Office of Naval

Research YIP Award under Grant No. N00014-16-1-2540. SM acknowledges financial support from Graduate School Summer Research Fellowship and the support from the FabLab at the University of Maryland Nanocenter. KJP acknowledges financial support from a National Defense Science and Engineering Graduate Fellowship.